\documentclass{article}
\pdfoutput=1
\usepackage{amsmath,amssymb,gensymb,color,graphicx,array,authblk,multirow,soul}
\usepackage{natbib,hyperref,aas_macros}
\usepackage[margin=22mm]{geometry}
\usepackage{listings}
\usepackage[utf8]{inputenc}
\newcommand{\code}[1]{\texttt{#1}}

\DeclareFixedFont{\ttb}{T1}{txtt}{bx}{n}{9} 
\DeclareFixedFont{\ttm}{T1}{txtt}{m}{n}{9}  
\DeclareRobustCommand{\hlblack}[1]{{\sethlcolor{black}\hl{#1}}}
\definecolor{deepblue}{rgb}{0,0,0.5}
\definecolor{deepred}{rgb}{0.6,0,0}
\definecolor{deepgreen}{rgb}{0,0.5,0}
\definecolor{border}{rgb}{0.9,0.9,0.9}
\definecolor{orange}{rgb}{1.0,0.5,0.0}
\definecolor{purple}{rgb}{0.7,0.0,0.5}
\newcommand{\good}[1]{{\color{deepgreen}\textbf{#1}}}
\newcommand{\bad}[1]{{\color{red}\textbf{#1}}}
\newcommand{\terrible}[1]{{\color{red}\textbf{\hlblack{#1}}}}
\newcommand{\medi}[1]{{\color{orange}\textbf{#1}}}
\newcommand{\mixed}[1]{{\color{blue}\textbf{#1}}}
\newcommand{\neutral}[1]{#1}
\newcommand{\dfn}[1]{\textbf{#1}}
\newenvironment{lowtabmargin}{\setlength{\tabcolsep}{0.5mm}}{}
\newcommand{\img}[1]{\resizebox{\hsize}{!}{\includegraphics{#1}}}

\title{How to avoid X'es around point sources in maximum likelihood CMB maps}

\author[1]{Sigurd K. Næss}
\affil[1]{Center for Computational Astrophysics, Flatiron Institute}

\begin{document}
\maketitle
\begin{abstract}
	The maximum likelihood estimator for CMB map-making is optimal and unbiased as long as
	the data model is correct, but in practice it rarely is, with model errors including
	sub-pixel structure and instrumental problems like time-variable gain and pointing errors.
	In the presence of such errors, the solution is biased, with the local error in each pixel
	leaking outwards along the scanning pattern by a noise correlation length. The most important
	sources of such leakage are strong point sources, and for common scanning patterns the leakage
	manifests as an X around each such source. I discuss why this happens, and present several
	old and new methods for mitigating and/or eliminating this leakage, along with a small
	stand-alone TOD simulator and map-maker in Python that implements them.
\end{abstract}

\section{Introduction}
Cosmic Microwave Background (CMB) telescopes map the sky by scanning an array of detectors repeatedly across
the sky, resulting in a time-series of samples $d$. This process is usually
modeled via the linear equation \citep{tegmark/map/1997}
\begin{align}
d &= Pm + n \label{eq:model}
\end{align}
where $d$ is the vector of samples,
$m$ is a vector representing the pixels of the sky we are trying to reconstruct, and
$P$ is a response matrix that encodes where the telescope was pointing on the sky
when each sample was taken and how it responded to that. $n$ represents the
noise contribution to each sample, and is usually taken to be normal distributed
with some covariance $N$. The maximum likelihood solution to this equation is
\begin{align}
	\hat m &= (P^T N^{-1} P)^{-1} P^T N^{-1} d \equiv M_{PN} d \label{eq:mapmaking}
\end{align}
where $M_N$ is the map-making operator using the response matrix $P$ and the noise model $N$.
It's simple to see that this solution is unbiased by inserting our model for $d$:
\begin{align}
\langle \hat m \rangle &= (P^T N^{-1} P)^{-1}P^T N^{-1}(Pm + \overbrace{\langle n \rangle}^{0}) = m
\end{align}
Given this result it might be surprising to hear that most maximum-likelihood maps
made using eq.~\ref{eq:mapmaking} are biased to some extent, with the archetypical example
being linear artifacts extending away from bright point sources in the map, often in the shape of an X
(but this depends on the telescope's scanning pattern). The source of these artifacts is \emph{model
error}: the failure of eq.~\ref{eq:model} to accurately describe the real data; and
in this paper I will describe the main types of model error in maximum-likelihood mapmaking, simulate their
effect and investigate several old and new methods for mitigating or eliminating them.

\section{Sources of model error}
\subsection{Sub-pixel structure}
The most obvious problem with eq.~\ref{eq:model} is that it models the sky as a
finite vector of pixels, which will leave out any signal on scales smaller than the
pixel spacing. This is exacerbated by the canonical choice of a \emph{nearest-neighbor} response
model in $P$. Every CMB map-maker I am aware of uses nearest-neighbor, including destripers like
MADAM \citep{madam/2010} and SRoll \citep{planck/hfi/maps/2018}, filter+bin map-makers like those
used in SPT \citep{spt/maps/2011} and BICEP \citep{bicep2a/2014} or the maximum likelihood map-makers
used in ACT \citep{act-2017} and QUIET \citep{quiet-gal/2015}. In this model, the value of
each sample is simply given by the value of the
closest pixel to it, regardless of how far away from the pixel center it is. This results
in a response matrix with a very simple structure: For each row (representing a sample)
a single element will be 1 (representing the pixel hit by that sample), and all others
will be zero.

\begin{figure}[ph!]
	\centering
	\begin{tabular}{cc}
		\bf uncorrelated & \bf correlated \\
		\includegraphics[width=0.48\textwidth]{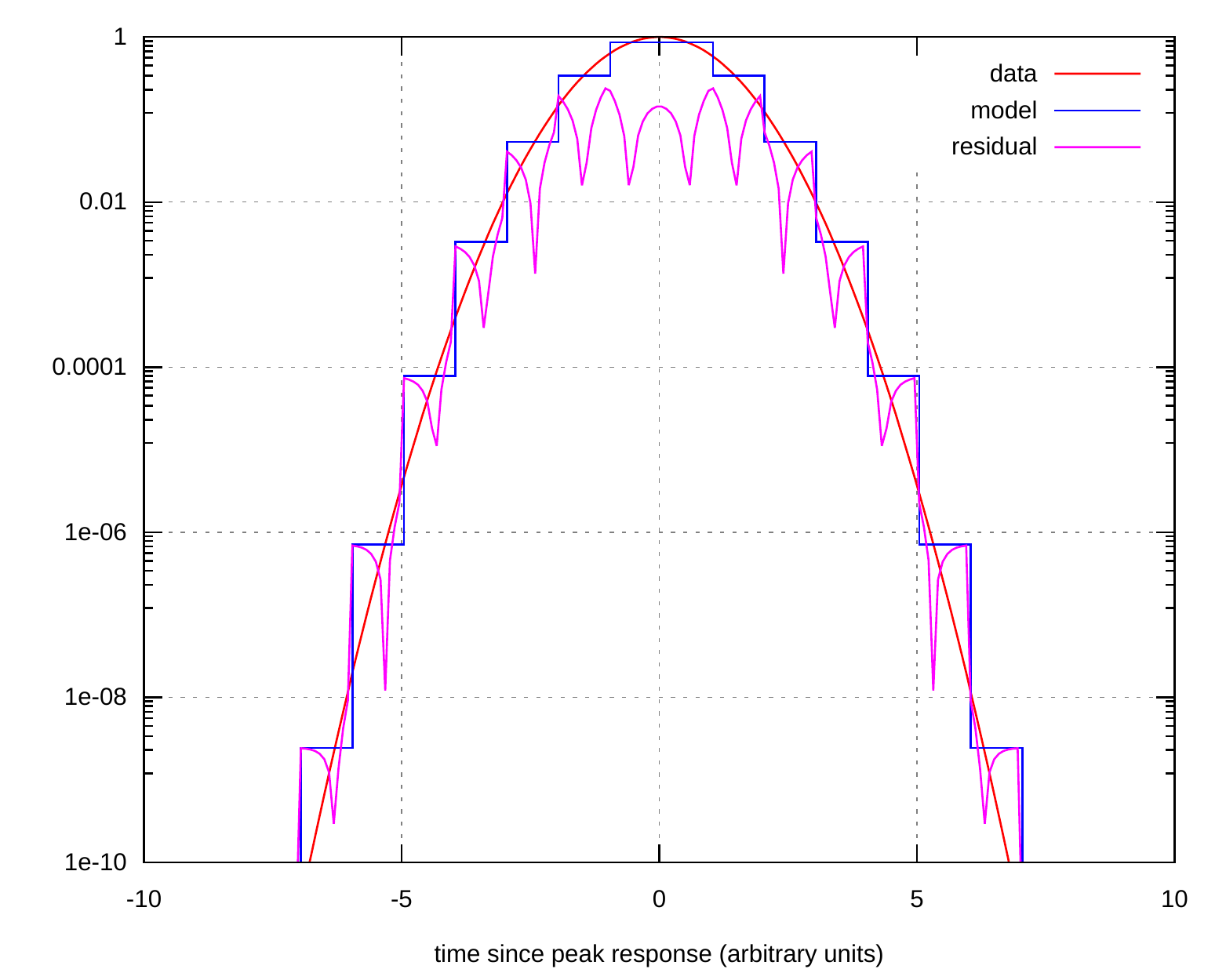} &
		\includegraphics[width=0.48\textwidth]{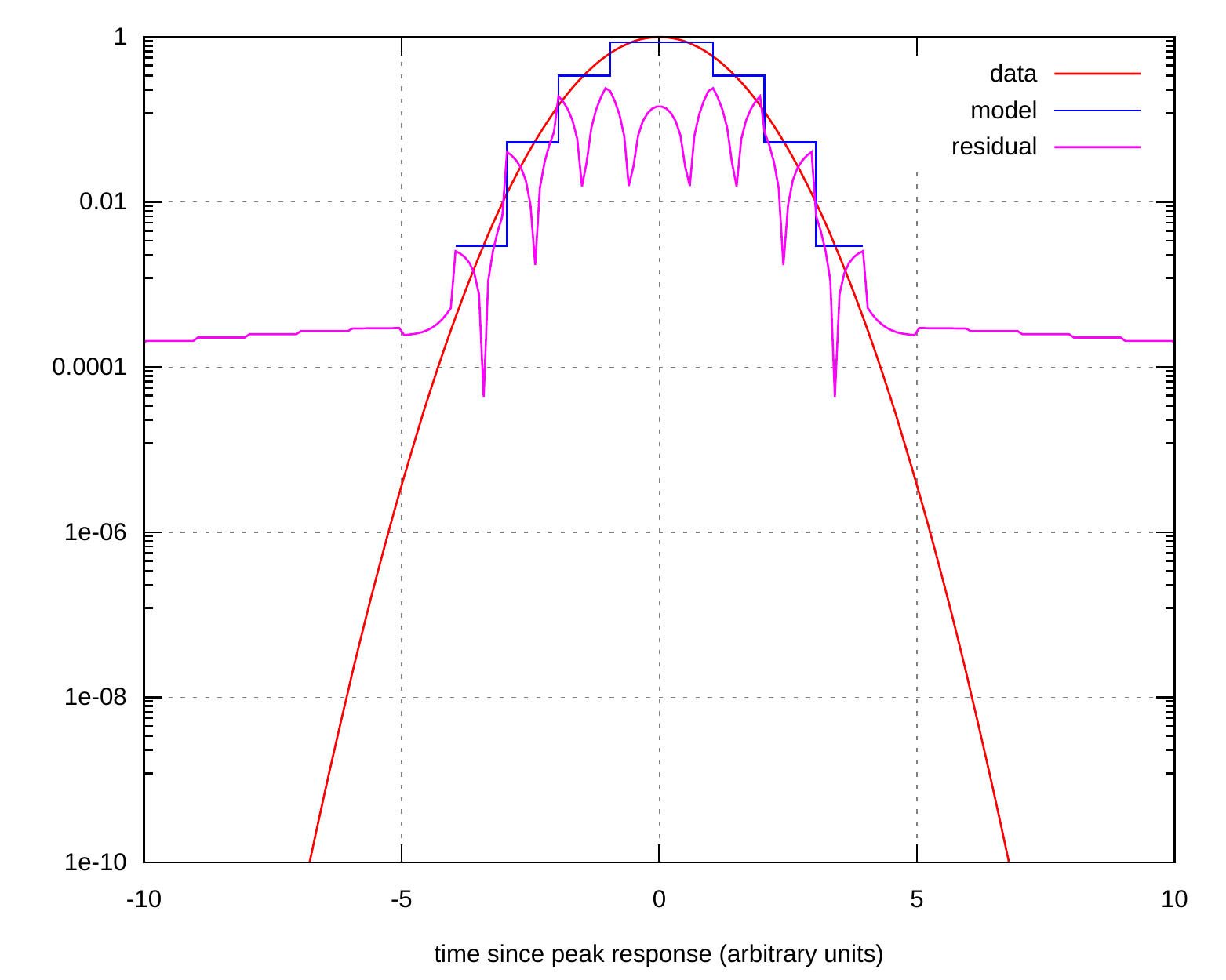}
	\end{tabular}
	\caption{A simulated data vector consisting of a noiseless scan through a smooth, Gaussian profile,
		compared to its nearest-neighbor-pixellated maximum likelihood
		model. \dfn{Left}: When an uncorrelated noise model is used ($N \propto I$)
		the residual is large but stays in the pixel where they were sourced. This error can
		be characterized as a simple pixel window. \dfn{Right}: When a correlated noise
		model is used the residuals are still large, but are now spread out over a larger area
		than the signal that sourced them.}
		\label{fig:subpix1d}
\end{figure}

Multiplication by a matrix with this structure is simple and efficient.
Given a map \code{m[ny,nx]} and arrays \code{y[nsamp]}, \code{x[nsamp]} containing the x and y pixel coordinates
of each sample, the forward operation $d = Pm$ can be implemented as
\begin{lstlisting}
for i in range(nsamp):
	d[i] = m[round(y[i]),round(x[i])]
\end{lstlisting}
and the transpose operation $m = P^Td$ can be implemented as
\begin{lstlisting}
for i in range(nsamp):
	m[round(y[i]),round(x[i])] += d[i]
\end{lstlisting}
However, this simplicity comes at a cost: The signal model implied by this
response matrix has a constant value inside each pixel and a discontinuous jump
at the edge of each pixel. This model and the resulting error is illustrated in figure~\ref{fig:subpix1d}.

A step-function model like this clearly cannot match the smooth behavior of the beam-convolved
sky, and the only term in eq. \ref{eq:model} that can absorb such a mismatch between the data
and the model is the noise term $n$. How this model error manifests is therefore determined by
our noise model $N$.

If $N$ is assumed to be uncorrelated ($N$ diagonal in time domain), then a high value of the residual
(honorary noise as far as our data model is concerned) in one sample does not tell us
anything about what the noise is doing in any other samples, and so the residual
just stays in the pixel that birthed it. The value in each pixel is therefore just
the mean of the samples that hit it (left panel of figure~\ref{fig:subpix1d}).
This is approximately
the mean value of the signal inside the area covered by the pixel, which is a straightforward
and useful thing to measure, and is not the sort of bias we are worried about here.

If, on the other hand, $N$ is assumed to be correlated, then the presence of high
``noise'' (actually residual) in some samples
will make the maximum likelihood prection for other samples within a noise correlation length
have a similar value for the noise. For long correlation lengths this could affect samples
many pixels away from the source of the residual. This is illustrated in the right panel
of figure~\ref{fig:subpix1d}. Here large sub-pixel residuals in the center result in a non-zero expectation
value for the noise at large distances. Given that the actual data has negligibly small values
there, the best fit model must cancel the expected noise, resulting in a non-zero model
extending far beyond the area with significant signal. Effectively the correlated
noise model is causing local model errors to leak roughly a correlation length
into the surrounding pixels.

\subsection{Signal or instrumental variability}
\label{sect:inconsistency}
Real telescopes don't have 100\% accurate pointing and gain, and even if they did, the
sky is not static and unchanging. These effects mean the telescope effectively sees the
sky jitter around slightly while varying in brightness. There is no room for this in
eq.~\ref{eq:model} aside from the noise term $n$, so time-variable instrumental errors
or variable point sources will lead to exactly the same sort of artifacts as sub-pixel errors do.

\section{Mitigation methods}
The artifacts are ultimately caused by a combination of two factors:
\begin{enumerate}
	\item Multiple samples are mapped onto the same pixel, but don't agree on what the
		signal in that pixel should be.
	\item The correlated noise model causes a non-local response to such errors.
\end{enumerate}
This points to two approaches for eliminating them: Improve the model so it matches
the data more accurately, or modify the noise model to make it more local.

\subsection{Higher-order mapmaking}
An obvious way to improve the model is to replace nearest-neibhor interpolation in 
the response matrix $P$
with smoother interpolation functions such as bilinear or bicubic spline interpolation.
These are popular in image processing, but are as far as I am aware unused in CMB map making.

Bilinear interpolation is considerably slower and more complicated than nearest neighbor,
and each sample now interacts with the four closest pixels instead of just one.
The forward operation $d=Pm$ is now (based on the implementaiton of \code{scipy.ndimage.map\_coordinates}):
\begin{lstlisting}
for i in range(nsamp):
	py1, px1 = floor(y[i]), floor(x[i])
	py2, px1 = py1+2, px1+1
	ry,  rx  = y[i]-py1, x[i]-px1
	vx1 = m[py1,px1] * (1-rx) + m[py1,px2] * rx
	vx2 = m[py2,px1] * (1-rx) + m[py2,px2] * rx
	d[i] = vx1 * (1-ry) + vx2 * ry
\end{lstlisting}
The transpose operation $m=P^Td$ is:
\begin{lstlisting}
for i in range(nsamp):
	py1, px1 = floor(y[i]), floor(x[i])
	py2, px1 = py1+2, px1+1
	ry,  rx  = y[i]-py1, x[i]-px1
	vx1 = d[i] * (1-ry)
	vx2 = d[i] * ry
	m[py1,px1] += vx1 * (1-rx)
	m[py1,px2] += vx1 * rx
	m[py2,px1] += vx2 * (1-rx)
	m[py2,px2] += vx2 * rx
\end{lstlisting}
Bicubic spline interpolation is another large step up in complexity. See the appendix for details.

Figure~\ref{fig:highorder1d} shows how these higher-order interpolation schemes perform for the same
1D Gaussian simulation we considered in figure~\ref{fig:subpix1d}
\begin{figure}[ph!]
	\centering
	\includegraphics[width=\textwidth]{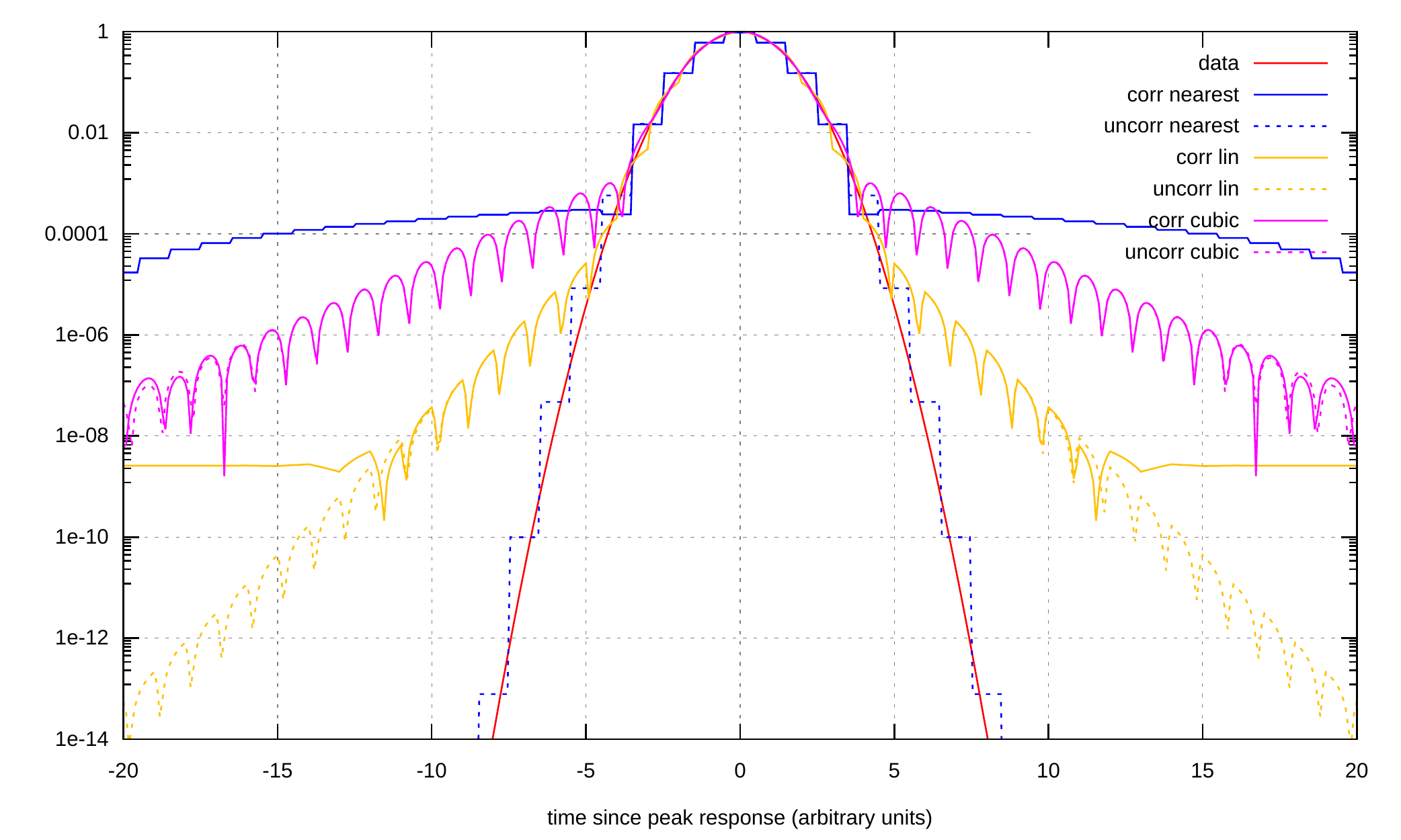}
	\caption{Comparison of the maximum-likelihood models for nearest neighbor,
		linear and cubic interpolation in the response matrix for uncorrelated and correlated
		noise models, for the same data as in figure~\ref{fig:subpix1d}. Linear and cubic interpolation fit the
		central part of the profile much better than nearest neighbor, leading to lower discrepancy
		between the data and model at high radius in the correlated case (solid curves).
		But this situation is reversed for the uncorrelated
		noise model (dashed curves). This happens because the linear and cubic models are inherently non-local,
		and are willing to sacrifice accuracy in areas where the signal is very weak in order to
		get a better match in areas with strong signal.}
	\label{fig:highorder1d}
\end{figure}
Both linear and cubic interpolation fit the peak of the Gaussian much better than nearest neighbor,
resulting in roughly 4 orders of magnitude lower spurious signal at high radius for the case with
a correlated noise model. However, they only reach this level at large radii. Before this, the
spurious signal is dominated by a new effect that was not present for nearest neighbor interpolation:
Higher order interpolation is inherently non-local, which means that each pixel is correlated with
its neighbors (and neighbors-neighbors for bicubic). This is responsible for the exponentially
decaying ringing pattern we see in the figure for these interpolation types, even for uncorrelated noise.
Effectively the fit is sacrificing accuracy at large radius in order to fit the strong, central
peak slightly better.

So we see that high interpolation order has smaller total residuals and hence lower
non-local artifacts from the noise correlations, but low-order interpolation has shorter
correlation length for the interpolation functions themselves, leading to a faster decay of
the ringing. For the example in figure~\ref{fig:highorder1d}, the sweet spot is linear interpolation.
Even better locality (at even higher cost) should be possible with an interpolation
function based on the actual point spread function of the instrument.

Aside from the high cost and non-locality inherent in these interpolators, they are also
limited by only addressing sub-pixel errors. Higher-order map-making does nothing to remove
artifacts from variable instrumental errors or intrinsic sky variability.

\subsubsection{Iterative approximation}
Since the response matrix often is the bottleneck in maximum-likelihood mapmaking,
it may be too expensive to make this step several times slower by using higher-order
mapmaking. A simple approximation that avoids most of this slowdown is to first make
a normal map using the fast, nearest-neighbor response matrix, and then subtract this
estimate of the sky from the time-ordered data using a bilinear or bicubic spline
response matrix. This process can be repeated if necessary, though in my tests there
was little point in doing more than 1-3 iterations.
\begin{align}
\hat m_0 &= 0 \notag \\
\hat m_{i+1} &= \hat m_i + M_{PN}(d - \bar P \hat m_i) \label{eq:iterative}
\end{align}
Here $P$ is the standard nearest-neighbor response matrix and $\bar P$ is
the higher-order one we are approximating. Note that this iteration converges
on an \emph{approximation} to the higher-order map, not the exact solution
we would have gotten by using $\bar P$ in the equation \ref{eq:mapmaking}.
However, as we shall see in the results section it does a pretty good job.

If one is already doing iterative mapmaking to avoid bias from
estimating the noise model $N$ from the full data $d$ rather than
just the noise $n$, then this iterative approximation to higher-order mapmaking
can simply be done as a part of that, and will incur almost no extra cost.

Alternatively, if one is solving for each map using an iterative method
like preconditioned conjugate gradients (CG), then the small scales that are the
main source of sub-pixel error often converge much faster than the larger ones.
In this case only a few CG steps are needed for all but the last iteration
of equation \ref{eq:iterative}, again resulting in almost no overhead compared to
standard mapmaking.

\subsection{Point source subtraction}
In figure~\ref{fig:subpix1d} we saw that non-local sub-pixel artifacts could leak far away from the
signal that sourced them, but it also bears noting that their amplitude was about
$10^4$ times lower than the source. For a smooth, low-contrast signal like the
cosmic microwave background, each pixel is contaminated by leakage from
other pixels in the vicinity, but since those other pixels have signal of similar
amplitude but random phase, one can expect each pixel to only be contaminated at the
$10^{-4}$ level, which is negligible. Non-local model errors are therefore only
a concern when a strong, compact signal is close to a region of weaker signal. In
CMB map-making the typical case would be a strong point source like a quasar.

If one knows the location and amplitude of the bright point sources in the sky,
one can simply subtract those point sources in the data before making the map. This
results in a map free of both the point sources (which can be useful in itself)
and the artifacts they would have sourced. If needed, the source model can then be
projected onto the sky using an uncorrelated noise model $W$ (for example $W=I$)
and added back to the map.
\begin{align}
\hat m_\textrm{srcfree} &= M_{PN}(d - \sum_i A_i B_i) \notag \\
\hat m_\textrm{tot} &= \hat m_\textrm{srcfree} + M_{PW}\sum_i A_i B_i
\end{align}
Here $A_i$ is the amplitude of source $i$, and $B_i$ is a vector of each sample's
response to an instrument beam centered on the position of that source.

This simple approach was used in e.g. \citet{dunner/etal/2013}, and succeeds in getting
rid of the artifacts sourced by sub-pixel structre in the point sources. However, unless
one has a point source model that takes into account time-variable pointing, gain or
beams errors, or intrinsic amplitude variations, then this method does nothing to
reduce artifacts from variability. It is also limiting that it requires some properties
of the sky to already be known, and that it can only be applied to signals for which
an accurate, non-pixellated model is available.

\subsection{White source mapping}
Model errors do not produce non-local artifacts when using an uncorrelated nosie model,
but of course, simply mapping using such a noise model would produce a horribly stripy
map in the presence of realistic correlated noise. However, we don't need everything to
be uncorrelated to get rid of leakage, we just need to decouple the bright point sources
from those correlations. This suggests the following approach:
\begin{enumerate}
	\item Make a point source free set of samples $d'$ from $d$ by smoothly in-painting the samples in $d$
		that hit the bright sources, ideally using maximum likelihood in-painting, but the quality of the
		in-painting only matters for the optimality of the method - bad in-painting will not cause a bias\footnote{
			This method splits the data into two parts, $d_\textrm{tot} = d_\textrm{inpaint} +
			(d_\textrm{tot}-d_\textrm{inpaint})$, and maps the first term with a correlated noise model
			and the last term with an uncorelated noise model. As long as both map-making methods
			are unbiased, the result will be unbiased regardless of the exact value of $d_\textrm{inpaint}$.
			In practice map-making with a correlated noise model can be biased due to model error
			artifacts, as we have seen. But for reasonable, smooth in-painting methods this bias
			will be completely negligible.}.
	\item Make a maximum-likelihood map $\hat m'$ of the in-painted TOD using the full noise model.
		\begin{align}
		\hat m' &= M_{PN} d'
		\end{align}
	\item If the in-painted regions were small compared to the noise correlation length, then
		the missing signal $d-d'$ from the previous map will have approximately white noise,
		so we can map these bothersome samples using a simple white noise model:
		\begin{align}
		\Delta m &= M_{PW}(d-d')
		\end{align}
	\item Add these maps to get the total, unbiased sky map: $\hat m = \hat m' + \Delta m$.
\end{enumerate}

\subsection{Source sub-sampling}
We can completely eliminate model error by having at least one degree of freedom per sample,
but if we did that to every sample, there would be no averaging down of the noise,
and we wouldn't really get a map, just a reorganized version of the input data. However,
there is nothing wrong with oversampling like this just around strong point sources and
other problematic regions, as long as these regions are small compared to the noise
correlation length. This results in the model
\begin{align}
d &= Pm + Gs + n \notag \\
	&= \begin{bmatrix}P&G\end{bmatrix}\begin{bmatrix}m&s\end{bmatrix}^T + n \label{eq:joint}
\end{align}
Here $s$ are the new per-sample degrees of freedom for the samples that hit bright sources,
and $G$ the matrix that maps them 1-to-1 only the corresponding samples. To be explicit,
$a=Gs$ could be implemented as \code{a[mask] = s}, and $b=G^Td$ as \code{b = d[mask]},
where mask is True for samples that hit bright sources and False otherwise.

One can then solve jointly for the maximum-likelihood values $\hat m$ and $\hat s$. This can still
be done using equation~\ref{eq:mapmaking}, since equation~\ref{eq:joint} is still of the same form
as equation~\ref{eq:model}. $m$ and $s$ are in isolation not well defined: Some samples are
both covered by pixels from $m$ and by special per-sample degrees of freedom from $m$, so
signal can move freely from one to the other without changing the residuals. This can be avoided by
modifying $P$ to have it skip those samples, but it is not actually necessary, for
the total map
\begin{align}
\hat m_\textrm{tot}&= \hat m + M_{PW}\hat s \label{eq:jointsol}
\end{align}
\emph{is} well-defined in either case, and is the artifact-free map of the sky we are after.

An advantage of this method is that it is very similar to a common way to handle
per-sample cuts in maximum-likelihood map-making. There one also solves for an extra degree
of freedom per cut sample. The only difference is that we here don't throw that part of the
solution away afterwards, but instead project them back into the final map using a white noise model
(see section~\ref{sect:source-cut}).

%

\subsection{Source cutting}
\label{sect:source-cut}
Perhaps the conceptually simplest way to avoid model errors from small problematic regions
of the sky is to simply exclude samples hitting those regions from the mapping process. In
practice, however, most correlated noise models are best represented in Fourier space, and fast Fourier
transforms cannot easily handle missing samples. Hence, simply dropping those samples isn't
straightforward. A trick that effectively achieves the same thing is to modify the
pointing matrix to give the bad samples a dummy degree of freedom each instead of pointing
them at the sky \citep{dunner/etal/2013},
\begin{align}
d &= \bar P m + Gs + n
\end{align}
Here $G$ has the same definition as in eq.~$\ref{eq:joint}$, selecting only the sources hitting
the bright objects, and $\bar P$ is identical to $P$ except that those samples are skipped.
The solution to this is equivalent to $\hat m_\textrm{tot}$ from eq.~\ref{eq:jointsol} with
the cut region masked. To see this, notice that replacing $\bar P$ with $P$ will not change
the solution outside the cut area because $\bar P$ already has enough degrees of freedom to
perfectly match the data there, resulting in zero local residual, and adding even more degrees of freedom
in this area will not change that.

\subsection{PGLS}
Postprocessed Generalized Least Squares (PGLS) is a method developed for mapmaking with
high-S/N, high dynamic range data from the infrared telescope Herschel, where artifacts
are much more pronounced than in CMB maps \citep{pgls-2012}. Unlike the other methods discussed here, this method
does not try to prevent artifacts from forming, but instead attempts to measure and subtract
them in a second pass of map-making. We can model a standard maximum-likelihood map built
using eq.~\ref{eq:mapmaking} as in practice being $\hat m = m + a + q$ where $m$ is the
map we would have gotten for a noise-free dataset with an uncorrelated noise model,
$a$ are the unwanted artifacts and $q$ is the pixel-space noise.
The goal of PGLS is build an estimator $\hat a$ for the artifacts in the map, and then subtract
this from $\hat m$ to produce a clean map.

We can exploit the fact that $a$ is a locally poor fit to the data to isolate it from the signal.
Projecting back into time domain we get 
\begin{align}
	r &= P\hat m - d = Pm - d + Pa + Pq
\end{align}
Ideally the term $d-Pm$ would equal our time-domain noise $n$, but as we have seen it
also contains sub-pixel errors $e$ (which is why the artifacts arise in the first place).
Hence, we have
\begin{align}
	r &= Pa + Pq - n - e
\end{align}
The sub-pixel errors average to zero when averaged over a pixel using an uncorrelated noise model\footnote{
	This follows from our definition of $m$ above. $M_{PW}e = M_{PW}(s - PM_{PW}s) = 0$, where $s$ is the noise-free sky signal.}, so if we could get rid of the
correlated noise in the term $Pq-n$ we could recover $\hat a$ by mapping $r$ with an uncorrelated
noise model. PGLS does this by applying a running median filter $f$ to $r$, resulting in the following
algorithm:
\begin{align}
	\hat m_0 &= M_{PN} d \notag \\
	\hat m_i &= \hat m_{i-1} - M_{PW} f(P\hat m_{i-1} - d)
\end{align}
The median filter length is a compromise between the need to remove as much correlated noise as
possible and to retain the artifacts themselves, and must be tuned to the instrument. The balance
between artifacts and noise can also be adjusted by iterating the scheme, as indicated by the
index $i$. Each iteration recovers more of the artifacts at the cost of letting through more noise.
For the toy example I test in this paper I found a median filter width of 21 samples and 5 iterations
to work well.

\subsection{XGLS}
The reason why local errors in areas of high signal end up leaking into the surroundings
is ultimately that the noise model does not expect these areas to be more prone to high
``noise'' than other regions. We can remedy this if we have an estimate of sub-pixel variance
by addin this as an extra term $X$ in the noise model such that $N_\textrm{tot} = 
N + X$, where $X_{ij} = \delta_{ij} \sigma_i^2$ and where $\sigma_i$ is the standard deviation
of the sub-pixel signal variability in the pixel hit by sample $i$. Given this noise model one can
produce an optimal, artifact-free map as $\hat m = M_{PN_\textrm{tot}} d$. This is the
Pixel Noise Generalized Least Squares (XGLS) estimator \citep{xgls-2017}, and like PGLS it
was developed for high S/N, high-dynamic range infrared data from the Herschel space telescope.

In many ways this is the proper way to solve the problem - instead of more or
less ad-hoc workarounds one tackes the problem at its root cause by telling the
noise model about this extra source of variance. This elegance comes at a cost, though:
\begin{enumerate}
	\item One needs an estimate of the sub-pixel signal variance. If one knew the noise-free
		signal $s$, then this could be estimated as $X = \textrm{diag}(PM_{PW}(s-PM_{PW}s)^2)$.
		Here $\textrm{diag}(x)$ indicates the diagonal matrix with $x$ on the diagoanal, and
		the squaring operation is performed sample-wise.
		For simplicity this is what I do in the simulations in this paper, but for real data
		where $s$ is unknown approximations are needed - see \citet{xgls-2017} for a discussion.
	\item While the sample-diagonal $X$ and the (usually) Fourier-diagonal $N$ are both easy
		to invert on their own, their sum is not diagonal in any simple basis and is therefore
		much harder to invert, typically requiring an iterative method like conjugate gradients (CG)
		Eq.~\ref{eq:mapmaking} itself is also usually solved with CG, resulting
		in a solution process where a full sequence of CG steps for the noise matrix inversion
		must be completed for every single CG step for the map. This makes XGLS dramatically
		slower than any of the other methods I tested. This is confounded by the outer CG iteration's
		sensitivity to inaccuracy in the inversion in the inner CG, necessitating a very strict
		convergence criterion there, further slowing things down. \citet{xgls-2017} report that
		they made this reasonably performant with some tuning, but still found the starting point
		of the iteration to have large effect on the solution. This is an indicator of convergence
		problems.
\end{enumerate}

\subsection{Summary of the methods}
Table~\ref{tab:summary} shows a summary of the methods we will investigate. The ideal method would be
blind (no prior knowledge about which parts of the sky have strong signal),
and would handle both sub-pixel errors and inconsistency at low cost, but none of the methods
tick all those boxes, and in particular none of the blind methods can handle inconsistency.

\begin{table}
	\centering
	\hspace*{-5mm}
	\begin{tabular}{cccccccccccc}
			&
			\bf std &
			\bf it.lin &
			\bf it.cubic &
			\bf bilin &
			\bf bicubic &
			\bf pgls &
			\bf xgls &
			\bf srcsub &
			\bf srcmask &
			\bf srcwhite &
			\bf srcsamp
		\\
			\bf blind &
			\good{yes} &
			\good{yes} &
			\good{yes} &
			\good{yes} &
			\good{yes} &
			\good{yes} &
			\good{yes} &
			\bad{no} &
			\bad{no} &
			\bad{no} &
			\bad{no}
		\\
			\bf subpix & 
			\bad{no} &
			\good{yes} &
			\good{yes} &
			\good{yes} &
			\good{yes} &
			\good{yes} &
			\good{yes} &
			\good{yes} &
			\good{yes} &
			\good{yes} &
			\good{yes}
		\\
			\bf incon. &
			\bad{no} &
			\bad{no} &
			\bad{no} &
			\bad{no} &
			\bad{no} &
			\bad{no} &
			\good{yes} &
			\bad{no} &
			\good{yes} &
			\good{yes} &
			\good{yes}
		\\
			\bf holes &
			\good{no} &
			\good{no} &
			\good{no} &
			\good{no} &
			\good{no} &
			\good{no} &
			\good{no} &
			\good{no} &
			\bad{yes} &
			\good{no} &
			\good{no}
		\\
			\bf cost &
			\good{1} &
			\good{1*} &
			\good{1*} &
			\bad{4} &
			\bad{16} &
			\good{1} &
			\terrible{500} &
			\good{1} &
			\good{1} &
			\good{1} &
			\good{1}
		\\
			\bf compl. &
			\good{v.low} &
			\medi{high} &
			\bad{v.high} &
			\medi{high} &
			\bad{v.high} &
			\good{low} &
			\medi{high} &
			\good{low} &
			\good{low} &
			\good{low} &
			\good{low}
		\\
			\bf pixwin &
			\neutral{std} &
			\mixed{small} &
			\mixed{small} &
			\mixed{small} &
			\mixed{small} &
			\neutral{std} &
			\neutral{std} &
			\neutral{std} &
			\neutral{std} &
			\neutral{std} &
			\neutral{std}
		\\
			\bf new &
			\neutral{no} &
			\mixed{yes} &
			\mixed{yes} &
			\mixed{yes} &
			\mixed{yes} &
			\neutral{no} &
			\neutral{no} &
			\neutral{no} &
			\neutral{no} &
			\mixed{yes} &
			\mixed{yes}
	\end{tabular}
	\caption{Summary of the methods discussed in this paper. The rows are: \dfn{blind}: Whether
		the method can be used on the whole map with no prior knowledge of which regions have
		large model errors; \dfn{subpix}: Whether the method targets sub-pixel errors;
		\dfn{incon.}: Whether the method targets inconsistency errors like gain and pointing
		errors; \dfn{cost}: A rough indication of time cost of the method, relative to
		standard map-making. Iterative linear/cubic mapmaking naively have a cost factor of
		$N_\textrm{iter}$, but in practice this can usually be avoided. \dfn{compl.}:
		A rough indication of how complex the method is to implement (this is somewhat subjective).
		\dfn{pixwin}:
		The pixel window the method produces. Higher-order interpolation methods results in
		a ``smaller'' (closer to unity) pixel window.\dfn{new}: Whether this is a new
		method or not. Those marked ``yes'' are, to my knowledge, not previously published.}
	\label{tab:summary}
\end{table}

\section{Simulations}
I wrote a minimal maximum-likelihood
mapmaking library \code{toy.py} in about 300 lines of Python to demonstrate the performance of these
methods, along with a driver script \code{examples.py}
that generates the plots in this paper. They can be found at \url{https://github.com/amaurea/model_error}.
The library emphasizes simplicity and is neither general
nor optimized for speed, and only depends on \code{numpy} and \code{scipy} (with
the exception of bilinear and bicubic map-making which also depends on \code{pixell}.
That said, the techniques described here have also been implemented in
the full Atacama Cosmology Telescope (ACT) mapmaker, and despite the simplicity of these test cases
the results are representative of running a real map-maker on real data.

\subsection{2D simulations}
The main simulations use two sets of 243 x 243 equi-spaced samples in a square grid,
the first ordered vertically and the second horizontally (the ordering matters for the
noise correlation structure). These are mapped onto an 81 x 81
pixel grid so that there are 3 x 3 samples per pixel per set. Using this setup I simulated
two signals:
\begin{enumerate}
	\item A Gaussian point source with a beam standard deviation of 1 pixel and an
amplitude of $2\cdot 10^4$ (500 times higher than the color scale used in the plots in the
results section). This will be representative of e.g. a bright quasar as seen by a high-resolution
CMB telesope like ACT.
	\item A CMB-like field built by simulating a $1/l$ spectrum with an angular
resolution of 1 pixel and an RMS of 1. This will test whether model error
is a problem in the absence of bright sources.
\end{enumerate}
I simulated two types of noise model:
\begin{enumerate}
	\item An uncorrelated noise model $N_\textrm{uncorr} = I$, representing the baseline case
		where no leakage is expected.
	\item A correlated noise model $N_\textrm{corr}(f_1,f_2) = \sigma^2 (1 + (f_1/f_\textrm{knee})^\alpha) \delta_{f_1,f_2}$. This is diagonal in Fourier space. $\alpha = -4$ represents an atmosphere-like steeply falling spectrum, meeting a noise floor with RMS $\sigma=1$ at the knee frequency $f_\textrm{knee} = 0.02$ in pixel units.
\end{enumerate}
Becuase it is the non-local weighting implied by a correlated noise model that causes leakage,
not the noise itself, one does not actually have to include the noise in the simulations.
To maximize S/N in the figures most of the simulations are noise-free, but I also include some noisy
simulations to show how effectively each method suppresses noise.

I simulated 3 types of model errors:
\begin{enumerate}
	\item Sub-pixel signal was simulated generating the signals at higher resolution than the
		output maps, as described above. This was included in every simulation.
	\item Gain errors were simulated by decreasing/increasing the signal amplitude by
	0.5\% for the vertical/horizontal data set, representing a total gain mismatch of
	1\%. This was included in the simulations labeled ``gain error''.
	\item Pointing errors were simulated by offsetting the horizontal dataset by 0.01
		pixels diagonally relative to the vertical dataset. This was includes in the
	simulations labeled ``pt. error''.
\end{enumerate}

\subsection{1D simulations}
I also produced as smaller number of 1-dimensional simulations for figure~\ref{fig:subpix1d} and figure~\ref{fig:highorder1d}.
These used the same noise models and point source signal as the 2-dimensional simulations,
but consisted of just a single scan of length 1100 samples which was mapped onto 100 pixels.

\section{Results}
\begin{figure}[ph!]
	\centering
	\begin{lowtabmargin}
	\begin{tabular}{>{\centering\arraybackslash}m{15mm}>{\centering\arraybackslash}m{12mm}>{\centering\arraybackslash}m{12mm}>{\centering\arraybackslash}m{12mm}>{\centering\arraybackslash}m{12mm}>{\centering\arraybackslash}m{12mm}>{\centering\arraybackslash}m{12mm}>{\centering\arraybackslash}m{12mm}>{\centering\arraybackslash}m{12mm}>{\centering\arraybackslash}m{12mm}>{\centering\arraybackslash}m{12mm}>{\centering\arraybackslash}m{12mm}>{\centering\arraybackslash}m{12mm}}
		 &
		\bf \footnotesize standard &
		\bf \footnotesize standard &
		\bf \footnotesize it.lin &
		\bf \footnotesize it.cubic &
		\bf \footnotesize bilin &
		\bf \footnotesize bicubic &
		\bf \footnotesize pgls &
		\bf \footnotesize xgls &
		\bf \footnotesize srcsub &
		\bf \footnotesize srccut &
		\bf \footnotesize srcwhite &
		\bf \footnotesize srcsamp
	\\
		&
		\footnotesize uncorr &
		\footnotesize corr &
		\footnotesize corr &
		\footnotesize corr &
		\footnotesize corr &
		\footnotesize corr &
		\footnotesize corr &
		\footnotesize corr &
		\footnotesize corr &
		\footnotesize corr &
		\footnotesize corr &
		\footnotesize corr
	\\
		\bf \footnotesize plain &
		\img{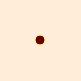} &
		\img{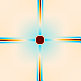} &
		\img{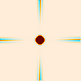} &
		\img{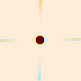} &
		\img{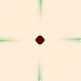} &
		\img{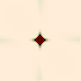} &
		\img{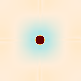} &
		\img{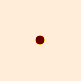} &
		\img{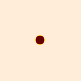} &
		\img{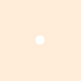} &
		\img{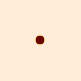} &
		\img{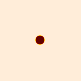}
	\\
		\bf \footnotesize gain error &
		\img{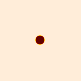} &
		\img{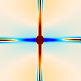} &
		\img{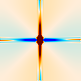} &
		\img{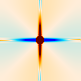} &
		\img{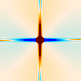} &
		\img{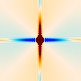} &
		\img{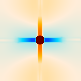} &
		\img{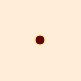} &
		\img{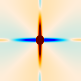} &
		\img{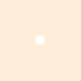} &
		\img{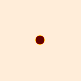} &
		\img{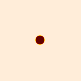}
	\\
		\bf \footnotesize pt. error &
		\img{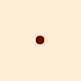} &
		\img{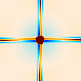} &
		\img{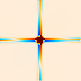} &
		\img{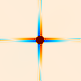} &
		\img{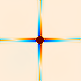} &
		\img{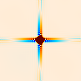} &
		\img{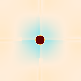} &
		\img{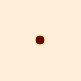} &
		\img{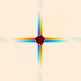} &
		\img{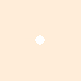} &
		\img{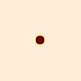} &
		\img{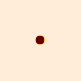}
	\\
		\bf \footnotesize plain + noise &
		\img{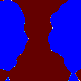} &
		\img{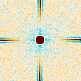} &
		\img{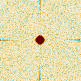} &
		\img{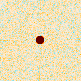} &
		\img{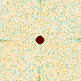} &
		\img{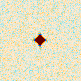} &
		\img{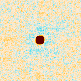} &
		\img{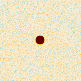} &
		\img{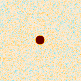} &
		\img{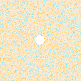} &
		\img{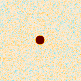} &
		\img{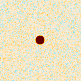}
	\\
		\bf \footnotesize gain error + noise &
		\img{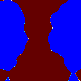} &
		\img{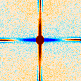} &
		\img{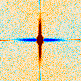} &
		\img{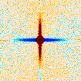} &
		\img{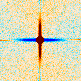} &
		\img{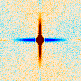} &
		\img{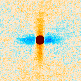} &
		\img{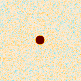} &
		\img{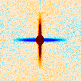} &
		\img{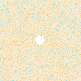} &
		\img{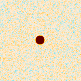} &
		\img{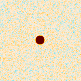}
	\\
		\bf \footnotesize pt. error + noise &
		\img{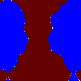} &
		\img{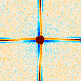} &
		\img{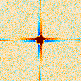} &
		\img{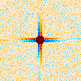} &
		\img{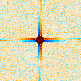} &
		\img{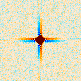} &
		\img{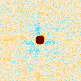} &
		\img{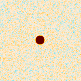} &
		\img{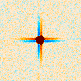} &
		\img{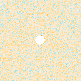} &
		\img{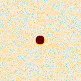} &
		\img{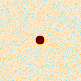}
	\\
		\bf \footnotesize plain + cmb + noise &
		\img{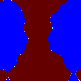} &
		\img{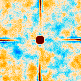} &
		\img{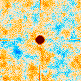} &
		\img{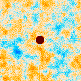} &
		\img{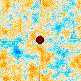} &
		\img{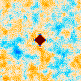} &
		\img{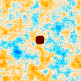} &
		\img{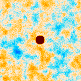} &
		\img{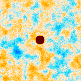} &
		\img{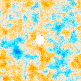} &
		\img{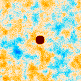} &
		\img{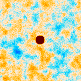}
	\\
		\bf \footnotesize gain error + cmb + noise &
		\img{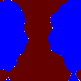} &
		\img{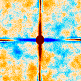} &
		\img{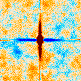} &
		\img{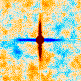} &
		\img{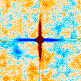} &
		\img{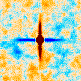} &
		\img{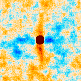} &
		\img{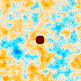} &
		\img{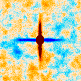} &
		\img{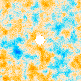} &
		\img{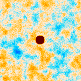} &
		\img{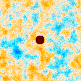}
	\\
		\bf \footnotesize pt. error + cmb + noise &
		\img{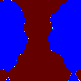} &
		\img{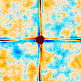} &
		\img{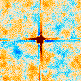} &
		\img{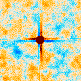} &
		\img{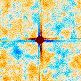} &
		\img{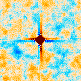} &
		\img{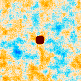} &
		\img{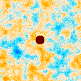} &
		\img{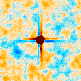} &
		\img{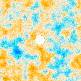} &
		\img{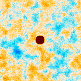} &
		\img{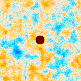}
	\end{tabular}
	\end{lowtabmargin}
	\caption{Model errors around a bright point source for the
		different methods discussed in this paper, for a simulation with 3x3
		samples uniformly distributed inside each pixel for each of a vertical and
		horizontal scanning pattern. The top third is for a noise free
		simulation, the second third includes strongly correlated noise,
		and the bottom third adds a CMB-like smooth signal.
		The rows are: \dfn{plain}: no special conditions; \dfn{gain error}:
		the vertical scanning pattern sees a 1\% higher signal than the horizontal
		scanning pattern; and \dfn{pt. error}: the horizontal scanning pattern
		sees the signal offset by 0.01 pixel diagonally. The column headers refer to
		the different mapmaking algorithms described in the text. The sub-headers
		indicate whether a correlated or uncorrelated noise model was used. The
		color range is $\pm 10^{-4}$ of the peak source amplitude.}
	\label{fig:srcleak}
\end{figure}

\begin{table}
	\centering
	\resizebox{\textwidth}{!}{%
	\begin{tabular}{cccccccccccc}
			&
			\bf standard &
			\bf it.lin &
			\bf it.cubic &
			\bf bilin &
			\bf bicubic &
			\bf pgls &
			\bf xgls &
			\bf srcsub &
			\bf srccut &
			\bf srcwhite &
			\bf srcsamp
		\\
			\bf near &
			\bad{0.868} &
			\medi{0.143} &
			\medi{0.270} &
			\medi{0.172} &
			\bad{1.777} &
			\medi{0.295} &
			\good{0.000} &
			\good{0.000} &
			\good{0.000} &
			\good{0.000} &
			\good{0.000}
		\\
			\bf far &
			\bad{3.518} &
			\bad{1.146} &
			\medi{0.392} &
			\bad{0.889} &
			\medi{0.178} &
			\medi{0.201} &
			\good{0.000} &
			\good{0.000} &
			\good{0.000} &
			\good{0.000} &
			\good{0.000}
		\\
			\bf near + pt. &
			\bad{2.544} &
			\bad{2.016} &
			\bad{1.965} &
			\bad{2.058} &
			\bad{3.614} &
			\medi{0.355} &
			\good{0.000} &
			\bad{1.909} &
			\good{0.000} &
			\good{0.000} &
			\good{0.000}
		\\
			\bf far + pt. &
			\bad{3.603} &
			\bad{1.241} &
			\bad{1.248} &
			\bad{1.212} &
			\bad{1.239} &
			\medi{0.257} &
			\good{0.000} &
			\bad{1.235} &
			\good{0.000} &
			\good{0.000} &
			\good{0.000}
		\\
			\bf near + gain &
			\bad{5.668} &
			\bad{4.593} &
			\bad{4.928} &
			\bad{4.605} &
			\bad{6.425} &
			\bad{2.065} &
			\good{0.000} &
			\bad{4.799} &
			\good{0.000} &
			\good{0.000} &
			\good{0.000}
		\\
			\bf far + gain &
			\bad{3.823} &
			\bad{2.896} &
			\bad{3.026} &
			\bad{2.877} &
			\bad{3.034} &
			\bad{1.488} &
			\good{0.000} &
			\bad{3.116} &
			\good{0.000} &
			\good{0.000} &
			\good{0.000}
	\end{tabular}
	}
	\caption{The maximum absolute value of the leakage from a point source
		in units of $10^{-4}$ of the peak amplitude, measured in two distance bins: \dfn{near}:
		5-10 pixels from from the center (just outside the point where the beam becomes negligible),
		and \dfn{far}: beyond 10 pixels from the center. The leakage is measured as the
		difference between a map using the given model and a correlated noise model and a map using
		standard nearest neighbor with a white noise model. For the beam, pixel size and noise model
		used here, sub-pixel errors result in O($10^{-4}$) leakage. Higher-order interpolation reduces this
		by a factor of 0.5-20 depending on the distance and method (yes, in the worst case it makes it
		2x worse near the source), while the targeted methods completely eliminate sub-pixel errors.
		Once gain or pointing errors are added, only source cutting, white source mapping and
		source sub-sampling provide any meaningful improvement (though this will depend on the
		size of the errors chosen).}
	\label{tab:srcleak}
\end{table}

Figure~\ref{fig:srcleak} shows maximum likelihood maps for each method for a strong point source.
In the noise free case (top 3rd of the figure) standard nearest-neighbor mapmaking
with an uncorrelated noise model (first column) results in no leakage as expected, whether in the
presence of normal sub-pixel signal, gain errors or pointing errors. However, in the presence
of realistic correlated noise (mid 3rd of the figure) this method is extremely suboptimal,
resulting in a map completely
dominated by large-scale noise.

Standard map-making with a correlated noise model (second column)
down-weights the noise correctly, but causes an X-shaped pattern of leakage around the
source, regardless of whether the noise is actually present. This only gets worse in
the presence of gain and pointing errors. The size of the leakage relative to
the peak source amplitude is shown in table~\ref{tab:srcleak}, and is typically of order $3\cdot 10^{-4}$ for
this choice of beam, pixel size and noise model. While
this is small relative to the source itself, it might be much brighter than any
surrounding CMB for a sufficiently bright source.

Full and iterative linear and bicubic interpolation all behave similarly.
In the absence of gain and pointing errors they provide a moderate leakage reduction,
typically by a factor of 3-10, but varying by method and distance from the source.
The computationally heaviest method, bicubic interpolation, provides the best reduction (about a factor
of 20) of these at large distance, but as we saw in figure~\ref{fig:highorder1d}, this comes at the cost of making
things worse near the source.
The iterative approximation to bicubic interpolation might be the best compromise
between low-range and short-range leakage, reducing both to $<0.4 \cdot 10^{-4}$.
However, since these methods only target sub-pixel
errors, it comes as no surprise that they do not help at all with data inconsistencies
like gain and pointing errors, where the X returns in full force.

Source subtraction completely eliminates the artifacts in the ideal case, but
performs no better than the blind interpolation methods in the presence of gain
and pointing errors. The other targeted methods fare better:
Source cutting avoids artifacts even for inconsistent data, at the cost of a hole in the map,
while white source mapping and source sub-sampling avoid both these problems,
eliminating artifacts even in the presence of inconsistent data while still providing
a map of the source.

\begin{figure}[ph!]
	\centering
	\begin{lowtabmargin}
	\begin{tabular}{>{\centering\arraybackslash}m{15mm}>{\centering\arraybackslash}m{12mm}>{\centering\arraybackslash}m{12mm}>{\centering\arraybackslash}m{12mm}>{\centering\arraybackslash}m{12mm}>{\centering\arraybackslash}m{12mm}>{\centering\arraybackslash}m{12mm}>{\centering\arraybackslash}m{12mm}>{\centering\arraybackslash}m{12mm}>{\centering\arraybackslash}m{12mm}>{\centering\arraybackslash}m{12mm}>{\centering\arraybackslash}m{12mm}>{\centering\arraybackslash}m{12mm}}
			&
			\bf \footnotesize standard &
			\bf \footnotesize standard &
			\bf \footnotesize it.lin &
			\bf \footnotesize it.cubic &
			\bf \footnotesize bilin &
			\bf \footnotesize bicubic &
			\bf \footnotesize pgls &
			\bf \footnotesize xgls &
			\bf \footnotesize srcsub &
			\bf \footnotesize srccut &
			\bf \footnotesize srcwhite &
			\bf \footnotesize srcsamp
		\\
			&
			\footnotesize uncorr &
			\footnotesize corr &
			\footnotesize corr &
			\footnotesize corr &
			\footnotesize corr &
			\footnotesize corr &
			\footnotesize corr &
			\footnotesize corr &
			\footnotesize corr &
			\footnotesize corr &
			\footnotesize corr &
			\footnotesize corr
		\\
			\bf \footnotesize plain &
			\img{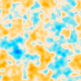} &
			\img{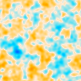} &
			\img{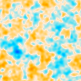} &
			\img{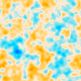} &
			\img{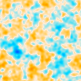} &
			\img{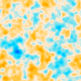} &
			\img{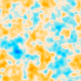} &
			\img{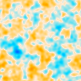} &
			\img{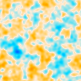} &
			\img{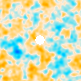} &
			\img{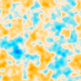} &
			\img{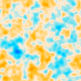}
		\\
			\bf \footnotesize gain error &
			\img{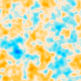} &
			\img{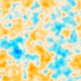} &
			\img{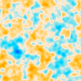} &
			\img{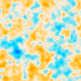} &
			\img{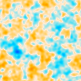} &
			\img{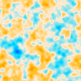} &
			\img{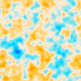} &
			\img{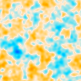} &
			\img{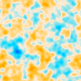} &
			\img{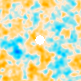} &
			\img{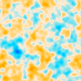} &
			\img{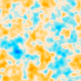}
		\\
			\bf \footnotesize pt. error &
			\img{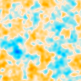} &
			\img{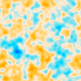} &
			\img{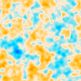} &
			\img{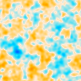} &
			\img{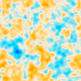} &
			\img{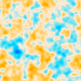} &
			\img{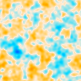} &
			\img{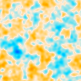} &
			\img{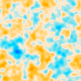} &
			\img{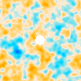} &
			\img{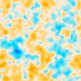} &
			\img{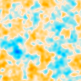}
		\\
			\bf \footnotesize plain resid &
			\img{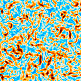} &
			\img{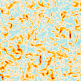} &
			\img{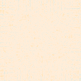} &
			\img{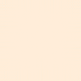} &
			\img{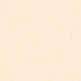} &
			\img{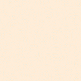} &
			\img{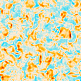} &
			\img{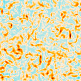} &
			\img{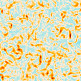} &
			\img{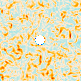} &
			\img{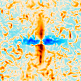} &
			\img{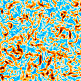}
		\\
			\bf \footnotesize gain error resid &
			\img{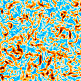} &
			\img{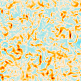} &
			\img{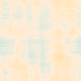} &
			\img{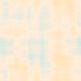} &
			\img{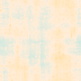} &
			\img{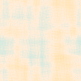} &
			\img{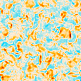} &
			\img{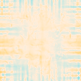} &
			\img{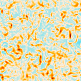} &
			\img{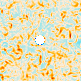} &
			\img{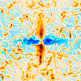} &
			\img{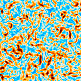}
		\\
			\bf \footnotesize pt. error resid &
			\img{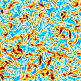} &
			\img{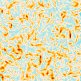} &
			\img{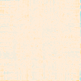} &
			\img{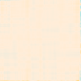} &
			\img{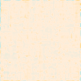} &
			\img{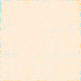} &
			\img{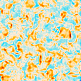} &
			\img{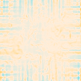} &
			\img{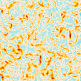} &
			\img{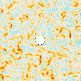} &
			\img{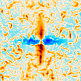} &
			\img{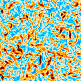}
		\\
			\bf \footnotesize plain wdiff &
			\img{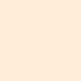} &
			\img{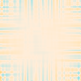} &
			\img{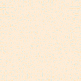} &
			\img{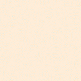} &
			\img{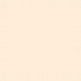} &
			\img{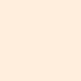} &
			\img{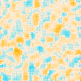} &
			\img{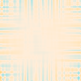} &
			\img{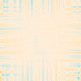} &
			\img{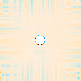} &
			\img{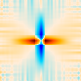} &
			\img{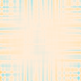}
		\\
			\bf \footnotesize gain error wdiff &
			\img{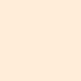} &
			\img{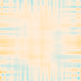} &
			\img{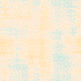} &
			\img{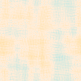} &
			\img{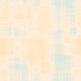} &
			\img{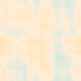} &
			\img{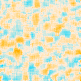} &
			\img{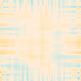} &
			\img{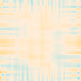} &
			\img{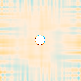} &
			\img{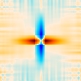} &
			\img{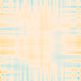}
		\\
			\bf \footnotesize pt. error wdiff &
			\img{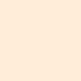} &
			\img{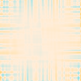} &
			\img{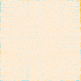} &
			\img{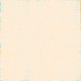} &
			\img{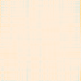} &
			\img{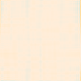} &
			\img{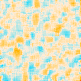} &
			\img{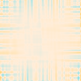} &
			\img{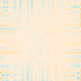} &
			\img{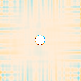} &
			\img{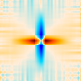} &
			\img{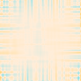}
	\end{tabular}
	\end{lowtabmargin}
	\caption{Like figure~\ref{fig:srcleak}, but for a noiseless CMB-like signal instead of
		a strong point source. \dfn{Top}: the resulting maps. \dfn{Middle}:
		the root-mean-square of the residual (data-model) per pixel. \dfn{Bottom}:
		The difference between each map and its uncorrelated version. The color
		range is $\pm 4$ in arbitrary units at the top and $\pm 0.1$ in the same units in
		the middle and bottom. The small gain and pointing errors we simulate here have
		almost no effect for low dynamic range fields like the CMB. The source-specific
		methods assume a zero-amplitude source in the patch center, even though none is
		actually present.}
	\label{fig:cmbleak}
\end{figure}

\begin{table}
	\centering
	\resizebox{\textwidth}{!}{%
	\begin{tabular}{cccccccccccc}
			&
			\bf standard &
			\bf it.lin &
			\bf it.cubic &
			\bf bilin &
			\bf bicubic &
			\bf pgls &
			\bf xgls &
			\bf srcsub &
			\bf srccut &
			\bf srcwhite &
			\bf srcsamp
		\\
			\bf plain &
			36.58 &
			21.48 &
			12.70 &
			\good{5.02} &
			\good{0.88} &
			\bad{150.84} &
			36.38 &
			36.58 &
			79.67 &
			\bad{274.90} &
			 36.82
		\\
			\bf pt. error &
			 38.68 &
			 33.55 &
			 26.29 &
			 14.35 &
			 13.78 &
			\bad{153.88} &
			 38.47 &
			 38.68 &
			 80.69 &
			\bad{275.80} &
			 38.97
		\\
			\bf gain error &
			 48.87 &
			 40.07 &
			 36.52 &
			 34.24 &
			 34.25 &
			\bad{152.00} &
			 48.70 &
			 48.87 &
			 80.69 &
			\bad{273.01} &
			 48.97
	\end{tabular}
	}
	\caption{The standard deviation of the difference between a map made with a correlated
		and uncorrelated noise model, in units of $10^{-4}$ of the standard deviation of the map itself, for each
		mapmaking method, and for a smooth, noiseless CMB-like field. The fractional leakage is typically
		O($4\cdot 10^{-3}$). This is larger than in table~\ref{tab:srcleak} because every point is a source of leakage, not just a single
		source, but since it is relative to the CMB instead of a bright point source, the amplitude is
		negligible in absolute terms. The white noise mapping method is an outlier, with relative
		error up to almost 3\% of the signal itself, but since this is a source-targeted method this would
		only happen near the targeted sources. Unlike table~\ref{tab:srcleak} the same mapmaking method
		is used for the maps being subtracted, to avoid penalizing methods with different pixel windows.}
	\label{tab:cmbleak}
\end{table}

Figure~\ref{fig:cmbleak} shows maximum-likelihood maps, residual RMS and deviation from the uncorrelated
solution for each method for a smooth, CMB-like field. It doesn't really make sense to apply
the source-targeted methods when there is no source present, but I still include them with
a dummy point source with zero amplitude in the patch center to see how these methods interact
with the surrounding CMB.

With a smooth, low-dynamic range field like this, the leakage is small
enough that the maps are practically identical. From table~\ref{tab:cmbleak} and the bottom
3rd of figure~\ref{fig:cmbleak} we see that the leakage is typically more than 200 times smaller than the
signal itself, even in the presence of gain and pointing errors.

The exception is white
source mapping, which has X-shaped leakage at 3\% of the signal strength.
This happens because the
maximum likelihood in-painting of the samples that hit the source is done independently
for the horizontal and vertical data set, which therefore end up disagreeing with
each other in this region. As we have seen before, such disagreement is interpreted
as noise, and spreads out by a noise correlation length. Like the X around strong
point sources, this too is a form of model error bias, but since it's sourced by
the CMB itself rather than a potentially very bright point source, it is never
more than a small fraction of the CMB. A 3\%-level contamination in a small fraction
of the sky is too small to matter in most circumstances, but to be safe one can
always use source sub-sampling instead, which doesn't have this issue.

\section{A note on polarization}
The toy example used for illustration in this article does not include polarization
in order to keep the code as simple and easy to understand as possible. That said,
the algorithms described here are general and straightforward to extend to polarization
(amounting mainly to a change in the implementation of the response matrix $P$), and
with the exception of PGLS and XGLS have been implemented successfully in the Atacama
Cosmology Telescope mapmaking framework \code{enlib}.

That said, it can be useful to consider the what the polarized properties of model error
artifacts are. The maximum-likelihood map $\hat m = M_{PN} d$ has a noise covariance
matrix $C = (P^TN^{-1}P)^{-1}$, and sub-pixel variance in some pixel $i$ source artifacts
$a_j \sim C^{\frac12}_{ij}$. That is: the artifacts leaking from a pixel are realizations
of the noise covariance structure around that pixel. We can use this to predict whether
artifacts will introduce significant I-to-P leakage. For a polarized map we can slice the
covariance matrix $C$ into Stokes and pixel dimensions: $C_{\alpha\beta ij}$, where
greek letters index Stokes parameters and latin one index pixels. Artifacts will have
significant I-to-P leakage if $C_{IP}$ is not very small relative to the
pure polarization components $C_{PP}$. Here $P$ is a stand-in for Q or U.

How large should we expect these terms to be a priori? In telescopes that measure
polarization via simultanous differencing, where pairs (or more) of detectors with different
polarization orientation measure the same spot on the sky simultanously, atmospheric
noise can be very effectively suppressed in polarization, leading to low $C_{IP}$.
The same is the case for telescopes with (rapidly) rotating half-wave plates or other polarization
modulators. This group of telescopes would have negligible I-to-P leakage in artifacts.
P-to-P artifacts - that is artifacts in polarization maps thate are sourced by
strong, local sources in polarization, like polarized point sources - would also be
greatly reduced because less correlated noise in polarization results in a short
correlation length and hence much more compact artifacts.

On the other hand, telescopes that rely on combining non-simultanous measurements
to separate the Stokes components would not suppress atmospheric noise in polarization.
An example of this would be a telescope with non-polarization-sensitive detectors that
uses a stepping half-wave plate for polarization separation. Here $C_{IP}$ would be
large, and artifacts sourced by unpolarized sources could end up being almost 100\%
polarized.

\section{Summary}
The model error mitigation strategies I have looked at can be grouped into four main classes:
\begin{enumerate}
	\item \dfn{Higher-order interpolation map-making}, which attempts to reduce sub-pixel errors by
		making the model interpolate smoothly between pixel centers instead of using a step function.
		These are blind, in that no prior knowledge of the sky is needed to apply them. They deal
		moderately well with sub-pixel errors, typically reducing them by a factor of 3-20
		in these tests
		(except very close to the source). They do not help at all with variability errors, however.
		As far as I am aware, these are new to the literature.
	\item \dfn{Source targeted methods}, which target a pre-defined set of regions of the sky, typically
		based on a point source database. With the exception of source subtraction, these can completely
		eliminate leakage from the targeted objects both from sub-pixel errors and from variability
		errors. Of these, source white mapping and source sub-sampling are new.
	\item \dfn{Artifact subtraction}, of which PGLS is the only example. It has the advantage
		of being a fast, blind method that handles both sub-pixel and variability errors moderately
		well, at the cost of some bias and increased correlated noise in the maps.
	\item \dfn{Signal-aware noise models}, of which XGLS is the only example. It is by far the
		highest quality blind method, at the cost of high complexity and extremely high
		computational cost.
\end{enumerate}
Given the high implementational complexity and limited improvement from higher-order map-making,
I do not think these methods are worth it, especially given the low impact of model errors outside
the neighborhood of strong point sources or similar high-contrast objects. XGLS performs much
better, tieing with the best source-targeted methods without needing any targeting itself. But its
cost is probably prohibitively high for realistic CMB datasets. Instead, I recommend
source targeted methods, and in particular the \dfn{source sub-sampling} method. Of the four source-targeted
methods I considered, it has the best performance, eliminating leakage even for variability errors
without leaving a hole in the map (as source cutting does) or introducing any secondary leakage
(as white source mapping does). It is also quite straightforward to apply, sharing most of its
implementation with an existing technique for handling per-sample data cuts. PGLS is also worth
considering if a blind method is needed, due to its simplicity and speed.

\section*{Acknowledgments}
The Atacama Cosmology Telescope data set provided the inspiration for this
paper. I would like to thank Jon Sievers and Reijo Keskitalo for useful discussion
about prior art. Flatiron Institute is supported by the Simons Foundation.

\bibliographystyle{act_titles}
\bibliography{refs}

\pagebreak

\appendix

\section{Bicubic spline interpolation}
The forward operation $d=Pm$ can be done in two parts. The first
is a convolution of the whole map by a spline pre-filter, followed by the
actual per-sample interpolation. This is implemented in \code{scipy.ndimage.map\_coordinates},
and when specializing to 2 dimensions and ignoring boundary conditions and optimization,
it can be written in pseudo-Python like this:
\begin{lstlisting}
def interpol(m, y, x):
	m = pre_filter(m)
	d = eval(m, y, x)
	return d

def pre_filter(m):
	m = m.copy()
	ny, nx = m.shape
	for py in range(ny): filter_1d(m[py,:])
	for px in range(nx): filter_1d(m[:,px])
	return m

def filter_1d(a):
	n = len(a)
	p = sqrt(3)-2
	for i in range(1,n):
		a[i] += p*a[i-1]
	a[n-1] *= -p/(1-p**2)
	for i in range(n-2,-1,-1):
		a[i] = p*(a[i+1]-a[i])

def eval(m, y, x):
	nsamp = len(y)
	for i in range(nsamp):
		py, px = floor(y[i]), floor(x[i])
		# offset of point from floored position
		ry, rx = y[i]-py, x[i]-px
		wy, wx = get_weights(ry), get_weights(rx)
		for jy in range(4):
			for jx in range(4):
				d[i] += wy[jy]*wx[jx]*m[py+jy-1,px+jx-1]
	return d

def get_weights(r):
	for j in range(4):
		# distance of each reference pixel from point
		d = abs(r+j-1)
		w[j] = (d*d*(d-2)*3+4)/6 if d < 1 else (2-d)**3/6
	return w
\end{lstlisting}
The transpose operation $m = P^T d$ is obtained by reversing the data flow:
\begin{lstlisting}
def interpol_trans(d, y, x):
	m = eval_trans(d, y, x)
	m = pre_filter_trans(m)
	return m

def pre_filter_trans(m):
	m = m.copy()
	ny, nx = m.shape
	for px in range(nx): filter_1d_trans(m[:,px])
	for py in range(ny): filter_1d_trans(m[py,:])
	return m

def filter_1d_trans(a):
	n = len(a)
	p = sqrt(3)-2
	a[0] *= -p
	for i in range(1, n-1):
		a[i] = p*(a[i-1]-a[i])
	a[n-1] -= a[n-2]
	a[n-1] *= -p/(1-p**2)
	for i in range(n-2,0,-1):
		a[i] += p*a[i+1]

def eval_trans(d, y, x):
	nsamp = len(y)
	m = zeros([ny,nx])
	for i in range(nsamp):
		py, px = floor(y[i]), floor(x[i])
		# offset of point from floored position
		ry, rx = y[i]-py, x[i]-px
		wy, wx = get_weights(ry), get_weights(rx)
		for jy in range(4):
			for jx in range(4):
				m[py+jy-1,px+jx-1] += wy[jy]*wx[jx]*d[i]
	return m
\end{lstlisting}
This transposed bicubic spline operation is missing from \code{sicpy}, but is
available in \code{pixell.interpol.map\_coordinates}. See its source
code for the full details (\url{https://github.com/simonsobs/pixell}).

\end{document}